\documentclass[fleqn,usenatbib]{mnras}

\usepackage{newtxtext,newtxmath}
\usepackage{xcolor}
\usepackage{hyperref}

\newcommand{\beq}{\begin{equation}}
\newcommand{\eeq}{\end{equation}}
\newcommand{\beqa}{\begin{eqnarray}}
\newcommand{\eeqa}{\end{eqnarray}}

\usepackage[T1]{fontenc}

\DeclareRobustCommand{\VAN}[3]{#2}
\let\VANthebibliography\thebibliography
\def\thebibliography{\DeclareRobustCommand{\VAN}[3]{##3}\VANthebibliography}

\usepackage{graphicx}	
\usepackage{amsmath}	

\title[The pcFLRW model and the time dependence of the Hubble constant]{The pseudo-complex Friedmann--Lema\^itre--Robertson--Walker model and the time 
dependence of the Hubble constant}

\author[L. Maghlaoui et al.]{
L. Maghlaoui,$^{1,2}$\thanks{E-mail: maghlaoui.th@gmail.com}
P. O. Hess,$^{3,4}$\thanks{E-mail: hess@nucleares.unam.mx}
F. Weber$^{5,6}$\thanks{E-mail: fweber@physics.ucsd.edu}
and C. A. Zen Vasconcellos$^{7,8}$\thanks{E-mail: cesaraugustozenvasconcellos@gmail.com}
\\
$^{1}$Physics Department, University of Mentouri, Constantine 1,
Constantine P. O. Box 325, Ain El Bey Way, 025017 Constantine, Algeria\\
$^{2}$Mathematical and Subatomic Physics Laboratory (LPMPS)
University of Mentouri, Constantine 1, Algeria\\
$^{3}$Instituto de Ciencias Nucleares, UNAM, Circuito Exterior, C.U.,
A.P. 70-543, 04510 M\'exico, D.F., Mexico\\
$^{4}$Frankfurt Institute for Advanced Studies,
J. W. von Goethe University, Hessen, Germany\\
$^{5}$Department of Physics, San Diego State University, SanDiego,CA92182,
United States of America\\
$^{6}$Department of Physics, University of California at SanDiego (UCSD),
La Jolla,CA92093, United States of America\\
$^{7}$International Center for Relativistic AstrophysicsNetwork (ICRANet),
Pescara 65122, Italy\\
$^{8}$Instituto de Física, Universidade Federal do Rio Grandedo Sul (UFRGS),
Porto Alegre 91501-970, Brazil
}

\date{Accepted XXX. Received YYY; in original form ZZZ}

\pubyear{2026}

\begin{document}
\label{firstpage}
\pagerange{\pageref{firstpage}--\pageref{lastpage}}
\maketitle

\begin{abstract}
The pseudo-complex version of the Friedmann--Lemaître--Robertson--Walker
model (pcFLRW) is presented within the framework of pseudo-complex
General Relativity (pcGR). In this approach, dark
energy arises as a
geometric consequence of the pseudo-complex structure, leading to a
time-dependent Hubble parameter rather than a strictly constant
$H_0$. The relation between the time derivative of the Hubble
parameter and a single geometric parameter $\beta$ in the effective
dark-energy equation of state is derived and
constrained using recent
DESI BAO data. Fitting $\beta$ yields a best-fit value $\beta = 1.0426
\pm 0.0144$, corresponding to a deceleration parameter $q = -0.9361
\pm 0.0216$ and a present-day Hubble
acceleration $\dot H_0 \simeq
(0.94 \pm 0.32)\times10^{-17}\,\mathrm{(km/s^2)/Mpc}$. Using
the exact Sandage–Loeb relation, the predicted redshift drift over 20
years for a source at $z=4$ is $\Delta v \simeq
-11.1\,\mathrm{cm\,s^{-1}}$, in     close agreement with the $\Lambda$CDM
prediction. 
In pcGR, however, the non-vanishing $\dot H_0$ is a direct geometric prediction, providing a clear and testable target for future high-precision spectroscopic observations.
\end{abstract}




\section{Introduction}
\label{intro}

The Hubble constant $H=\frac{{\dot a}}{a}$, where $a$ is the scale of
the universe and the dot its time derivative, describes the expansion
of the universe. Current measured values are $67.4\pm 0.5 ~ {({\rm
    km}/{\rm s})/{\rm Mpc}}$ from the Planck data
\citet{Planck2020,Hogg.1999}.  In contrast, local distance-ladder
measurements from Type Ia supernovae \citet{HubbleSNIa} calibrated with
Cepheid variables reported by the SH0ES collaboration, give a
systematic higher value of $73.30\pm1.04~ {\rm (km/s)/{\rm Mpc}}$
\citet{Riess2022}.  The persistent disagreement between the these two
independent measurements exceeds the $4\sigma$ level and is widely
referred to as the {\it Hubble tension}.

One objective of the DESI (Dark Energy Spectroscopic Instrument)
observation project \citet{DESI2024,DESI2025a,DESI2025b} is to measure
the distribution of the dark energy and also of the Hubble constant,
measuring the {\it Baryon Acoustic Oscillation} (BAO) and recent
results are available \citet{DESI-data}.  DESI aims to provide precise
measurements that may help resolve or clarify the Hubble tension.

In this contribution we present a particular model, the {\it
  pseudo-complex General Relativity} (pcGR) and its formulation of a
{\it Friedmann~-~Lema\^itre~-~Robertson-Walker model} (pcFLRW).  The
pcGR contains the contribution of dark energy, 
but instead of introducing a $\Lambda$ parameter, pcGR modifies the geometric foundation,
which {\it necessarily} introduces dynamics into the expansion history. The derived time
derivative of the Hubble constant ($\dot{H}_0$) is not an added parameter but a {\it consequence}
of this geometry.
Thus, while the Hubble tension in $\Lambda$CDM is fundamentally a crisis of constants, 
in pcGR it becomes a question of dynamics
(see for example \citet{Copeland}).
In contrast to $\Lambda$CDM, pcGR predicts a non-vanishing time
derivative of the Hubble parameter as a geometric consequence, making
DESI BAO measurements particularly appropriate for constraining the
model.
We emphasize that pcGR does not aim to resolve the Hubble tension by
shifting the value of $H_0$, but rather reframes it as a question of
cosmic dynamics through a non-vanishing $\dot H_0$.

The pcGR provides an equation of state for the dark energy, namely 
$p_\Lambda = - \beta \varrho_\Lambda$, where the index $\Lambda$ refers to the dark energy.
This $\beta$ value is fitted to the DESI data and knowing parameter value of $\beta$ 
the change in
time of the Hubble constant is deduced. 
Further observables are determined as the deceleration parameter  and the redshift drift. 

In section \ref{sec2} a short introduction to the pcGR is given, in a subsection the
pcFLRW model is explained and in section \ref{sec3} the results are compared to the DESI
observation. As a particular result we obtain an expression and numerical value of
the change for the Hubble constant in time.  
Also the redshift drift over 20 years is calculated.
Finally, in section \ref{sec4} Conclusions are drawn.

The metric signature used is $(-+++)$. The gravitational constant $G$ and the velocity of light $c$
are set to 1, 
except otherwise stated.

\section{A primer to pseudo-complex General Relativity (pcGR)}
\label{sec2}

The pcGR is presented in \citet{Hess2015} and 
in \citet{Hess2020} it is related to other modified GR models, it
is an algebraic extension of GR, i.e., the coordinate $x^\mu$
is extended to

\beqa
X^\mu & = & x^\mu + I y^\mu ~,~ I^2=1
~~,
\label{sec1-1}
\eeqa
which looks like a complex extension, except for $I^2=1$, which explains its name.
I is thus referred to as a pseudo-imaginary unit.
This implies an extension from a 4-dimensional space to an 8-dimensional one. As shown in
\citet{Kelly1986} this pc-extension is
the only algebraic extension permitted, because others include ghosts and/or tachyon 
solutions.

The length element is given by

\beqa
d\omega^2 & = & g_{\mu\nu}(X) dX^\mu dX^\nu
~~~,
\label{sec1-2}
\eeqa
where all elements are now pseudo-complex.

This metric and coordinates can be expressed within the so-called {\it zero-divisor basis}

\beqa
X^\mu & = & X^{\mu}_+ \sigma_+ + X^{\mu}_- \sigma_- ~,~ \sigma_\pm ~=~ 
\frac{1}{2} \left( 1 \pm I \right)
\nonumber \\
\sigma_\pm^2 & = & \sigma_\pm ~,~ \sigma_+\sigma_- ~=~ 0
\nonumber \\
g_{\mu\nu}(X) & = & g^+_{\mu\nu} (X_+) \sigma_+ + g^-_{\mu\nu} (X_-) \sigma_-
~~~.
\label{sec1-3}
\eeqa

Assuming the special case of a real metric, namely $g^{\pm}_{\mu\nu}(X_\pm) = g_{\mu\nu}(x)$,
the length element square acquires a simple form

\beqa
d\omega^2 & = & g_{\mu\nu}(x) \left[ dx^\mu dx^\nu + dy^\mu dy^\nu \right]
+ 2 I g_{\mu\nu}(x) dx^\mu dy^\nu
~~~.
\label{sec1-4}
\eeqa
Requiring that the length element is real leads to the condition

\beqa
g_{\mu\nu} dx^\mu dy^\nu & = & 0
~~~.
\label{sec1-5}
\eeqa

Here, we will discuss a model of the universe, which is known to be nearly flat, i.e.,
$g_{\mu\nu} \approx \eta_{\mu\nu}$, which is the Minkowski metric. As shown in \citet{Hess2020},
the solution is

\beqa
y^\mu & = & lu^\mu
~~~,
\label{sec1-6}
\eeqa
where for dimensional reasons a length parameter $l$ appears, which is interpreted as a minimal
length
(assumed to be the Planck length).
In this manner, the 8-dimensional space is reduced again to a 4-dimensional one.

A direct influence of the minimal length can be seen near
event horizons, predicting the appearance of a negative temperature and consequently an outward
pressure \citet{Leila2025}, contributing in such a manner to the stability of a black hole.

Note, that the $l$ is a {\it parameter} not affected by a Lorentz transformation. Thus, pcGR
permits a minimal length without violating the Lorentz/Poincar\'e symmetry. 
The advantages
of this property and practical/philosophical consequences are discussed in \citet{Weber2025}.
For more information, please consult the cited references.

In \citet{Hess2015} a variational principle was applied, which reads
$\delta S  ~\epsilon~ \sigma_-  A$. However, one can achieve the same by using
the standard variational principle, as it will be explained in what follows.

As an alternative, the standard variational principle was mentioned at the end 
of \citet{Hess2015} and explored in
\citet{Hess2020}, which uses the standard variational principle with a constraint
$g_{\mu\nu} dx^\mu dy^\nu = 0$ $\rightarrow$
$g_{\mu\nu} u^\mu u^\nu=C$, 
derived from (\ref{sec1-5}) and (\ref{sec1-6}),
namely the normalization constraint:

\beqa
\delta S & = & 0 ~,~ {\rm with}~ g^{\mu\nu}u_{\mu} u_{\nu} ~=~ C
\nonumber \\
S & = & \int dX^4 \sqrt{-g} \left( {\cal R} + 2\alpha \right) + S_m 
~~~,
\label{intro-2}
\eeqa
\citet{Hess2020}, where $S_m$ is the matter part.

This leads to an equation of motion of the form

\beqa
G_{\mu\nu} & = & 8\pi T^{\Lambda}_{\mu\nu} + 8\pi T^{m}_{\mu\nu}
~~~,
\label{intro-3}
\eeqa
where the label $\Lambda$ refers to dark energy and $T^{\Lambda}_{\mu\nu}$ 
is the $pc$-contribution.
Note, that $T^{\Lambda}_{\mu\nu}$ is the consequence of the implemented constraint and 
Eq. (\ref{sec1-6}), i.e. the minimal length is implicitly contained in the final result
through the dark energy fluid.

The variation with respect to $g^{\mu\nu}$ leads to the relation

\beqa
T^{\Lambda}_{\mu\nu} & = & \lambda u_\mu u_\nu + \alpha g_{\mu\nu}
~~~.
\label{intro-31}
\eeqa
This is the 
energy-momentum tensor
of an ideal fluid, where the Lagrange multiplier are given by
$\lambda = (p_\Lambda + \varrho_\Lambda )$ and $\alpha = -p$.

Consistent with the cosmological principle and the interpretation of dark energy as a homogeneous component, we model its energy-momentum tensor using the perfect fluid form:

\beqa
\left( T^{\Lambda}_{\mu\nu}\right) & = &
\left(
\begin{array}{cccc}
-\varepsilon_\Lambda & 0 & 0 & 0 \\
0 & p_\Lambda & 0 & 0 \\
0 & 0 & p_\Lambda & 0 \\
0 & 0 & 0 & p_\Lambda \\
\end{array}
\right)
~~~.
\label{intro-4}
\eeqa
The energy-momentum tensor $T^{\Lambda}_{\mu\nu}$ emerges from this variational principle with constraint, effectively encoding the geometric effects of the minimal length $\ell$ into an effective 
fluid description.

Of importance is the equation of state for the dark energy, given
by

\beqa
p_\Lambda & = & -\beta \varepsilon_\Lambda ~=~ w_\Lambda \varepsilon_\Lambda
~~~,
\label{intro-5}
\eeqa
where the $w_\Lambda$ is the effective dark energy equation of state parameter, usually used.

The parameter $\beta$ directly governs the evolution of the dark energy density. As we will show, deviations from $\beta = 1$ (corresponding to $\Lambda$CDM) lead to observable consequences in both the expansion history and the time evolution of the Hubble parameter.

\subsection{The pseudo-complex model of the Friedmann--Lema\^itre--Robert-son--Walker universe}

The pcFLRW model was first proposed in \citet{Leila2010} and resumed in
\citet{Hess2015}.  Here, we will present the main ingredients.

As explained in the last section, the length element $d\omega^2$ acquires the form
$d\omega^2$  =  $g_{\mu\nu} \left[ dx^\mu dx^\nu + dy^\mu dy^\nu \right]$.

In what follows, we exclude the second term $g_{\mu\nu} dy^\mu dy^\nu$ in the length element. This term is proportional to $\ell^2$ and is suppressed in the low-energy, large-scale limit relevant for cosmology. The dominant correction from the pseudo-complex extension is captured by the effective dark energy fluid (parameterized by $\beta$), which itself originates from the $\ell$-modified algebra integrated into the field equations.
Nevertheless, in future it would be interesting
to investigate the effects of an inclusion  of the second term in $d\omega^2$,
leading to higher order 
contributions.

From our theoretical model of the universe, we obtain a general form of the metric that is identical to that of the Standard Model.
However, the difference arises in the 
solution for the radius of the universe, $a(t)$. This modification also leads to an additional 
contribution in the Hubble constant. Furthermore, it introduces an effective equation of state for dark energy.
 
The pcFLRW metric is given by

\beqa
d\omega^2 & = & -dt^2 + a(t) \left( \frac{dr^2}{\left[1-kr^2\right]} + r^2 d\theta^2 
+r^2 {\rm sin}^2 (\phi ) d\phi^2 \right)
\label{FLRW-metric}
\eeqa

The pcFLRW equations are those from \citet{Hess2020,Leila2010} ($c=1$ and $G=1$) and are given by 

\beqa
8\pi \left( \rho + \varrho_\Lambda \right) & = &
\left[ \frac{3k}{a(t)^2} + 3 \frac{{\dot a} (t)^2}
{a(t)^2} \right]
\nonumber \\
8\pi\kappa \left(p+p_\Lambda \right) & = & 
-\left[ \frac{k}{a(t)^2} + \frac{{\dot a} (t)^2}
{a(t)^2} + \frac{2{\ddot a}(t)}{a(t)} \right]
~~~,
\label{eq11}
\eeqa
where we take $k=0$ for the flat universe.

The quantity $\varrho_\Lambda$ is the density of the dark energy and
$p_\Lambda$ the corresponding pressure. $p$ and $\varrho$ are the
pressure and density of either matter or radiation, or the sum of
them.

Here, we have also to specify 
the equation of state for the mass/radiation, which is defined as

\beqa
p & = & \alpha \varepsilon
~~~. 
\label{sec2-3}
\eeqa
The $\alpha$ has the value 0 for dust and $\frac{1}{3}$ for radiation, while $\beta$ is the central 
point of this contribution. 

As shown in \citet{Hess2015}, the
parameter $\beta$, which characterizes the dark energy equation of state in pcGR, is directly linked to the dynamics of the Hubble parameter via the relation:

\beqa
\beta  & = &  1 + \frac{2}{3}\frac{\dot{H}_0}{H_0^2}
~~~.
\label{sec2-4}
\eeqa
This equation reveals that a measurement of $\beta \neq 1$ implies a non-zero time derivative of the Hubble constant, $\dot{H}_0 \neq 0$, which can be compared to observation.

\section{ The DESI data and the time dependence of
  the Hubble constant within the pc-formulation}
\label{sec3}

The study of cosmological observables plays a fundamental role in
testing theoretical models of the Universe against high-precision
observational data. Among these observables, the Hubble
  parameter $H(z)$, the \textbf{comoving angular diameter distance}
$D_M(z)$, and \textbf{the volume-averaged distance} $D_V(z)$ are of
central importance.
In the standard $\Lambda$CDM framework, the Hubble parameter is strictly constant in time, implying $\dot H_0 = 0$ identically. In contrast, within pcGR a non-vanishing $\dot H_0 \neq 0$ arises as a direct and unavoidable consequence of the underlying pseudo-complex geometry, rather than from an additional phenomenological degree of freedom.
These quantities encode the expansion history of
the Universe and can be directly constrained by large-scale structure
surveys such as the Dark Energy Spectroscopic Instrument
  (DESI).

In the standard cosmological framework, based on General Relativity
(GR) and the $\Lambda$CDM model, the expansion rate is governed by the
equation,
\begin{equation}
H^2(z) = H_0^2 \left[ \Omega_m (1+z)^3 + \Omega_r (1+z)^4 + \Omega_\Lambda \right],
\end{equation}
where $H_0$ is the 
present-day Hubble constant
  while $H(z)$ denotes the redshift-dependent expansion rate.
The quantities
$\Omega_m$, $\Omega_r$, and $\Omega_\Lambda$ denote the present-day
density parameters of matter, radiation, and the cosmological
constant, respectively.

However, within the framework of pcGR, there are two important
modifications. First, a minimal length scale is introduced through the
pseudo-complex extension of the spacetime coordinates, as mentioned in
Section~\ref{sec2}. The second modification naturally leads to an effective
correction in the cosmological term, which can be characterized by a
dimensionless parameter $\beta$, as indicated in the following
equation~\citet{Hess2015,Leila2010}:

\begin{equation}
H^2(z) = H_0^2 \left[ \Omega_m (1+z)^3 + \Omega_r (1+z)^4 + \Omega_\Lambda (1+z)^{3(\beta - 1)} \right]
~~~,
\label{H-beta}
\end{equation}
where the exponent $3(\beta - 1)$ encodes deviations from the constant dark energy density predicted by standard $\Lambda CDM$ cosmology. 
The $\Omega_\Lambda$ term is a consequence of the inclusion of $y^\mu = l u^\mu$, i.e., of the minimal
length.

For $\beta = 1$, the model reduces to the standard case, while $\beta \neq 1$ represents the pcGR correction that can mimic evolving dark energy or modified gravity effects.
Inspecting Eq. (\ref{H-beta}), one notes that the dark energy is not constant but
depends on the parameter $\beta$, which depends dynamically on the
redshift $z$.

From this Hubble function, the key distance observables used in 
BAO (Baryon Acoustic Oscillation) analysis can be derived. The comoving distance $D_M(z)$ is defined by
(for comparison, we explicitly restore the speed of light $c \neq 1$):

\begin{equation}
D_M(z) = c \int_0^z \frac{dz'}{H(z')}
~~~,
\end{equation}
where $c$ is the speed of light. This quantity determines the transverse comoving separation between objects at redshift $z$.

The volume-averaged distance, often used in isotropic BAO analyses, combines both radial and transverse information:
\begin{equation}
D_V(z) = \left[ z^2 D_M^2(z)\frac{cz}{H(z)} \right]^{1/3}.
\end{equation}
Together, $H(z)$, $D_M(z)$, and $D_V(z)$ form the set of DESI
observables that can be directly compared with theoretical
predictions.  To perform this comparison, it is better to plot the
cosmological observables $H(z)$, $D_M(z)$, and $D_V(z)$ as functions
of the redshift $z$ for different values of $\beta$, using
(\ref{H-beta}).

\begin{figure}
   \begin{minipage}{0.48\textwidth}
     \centering
     \includegraphics[width=1\linewidth]{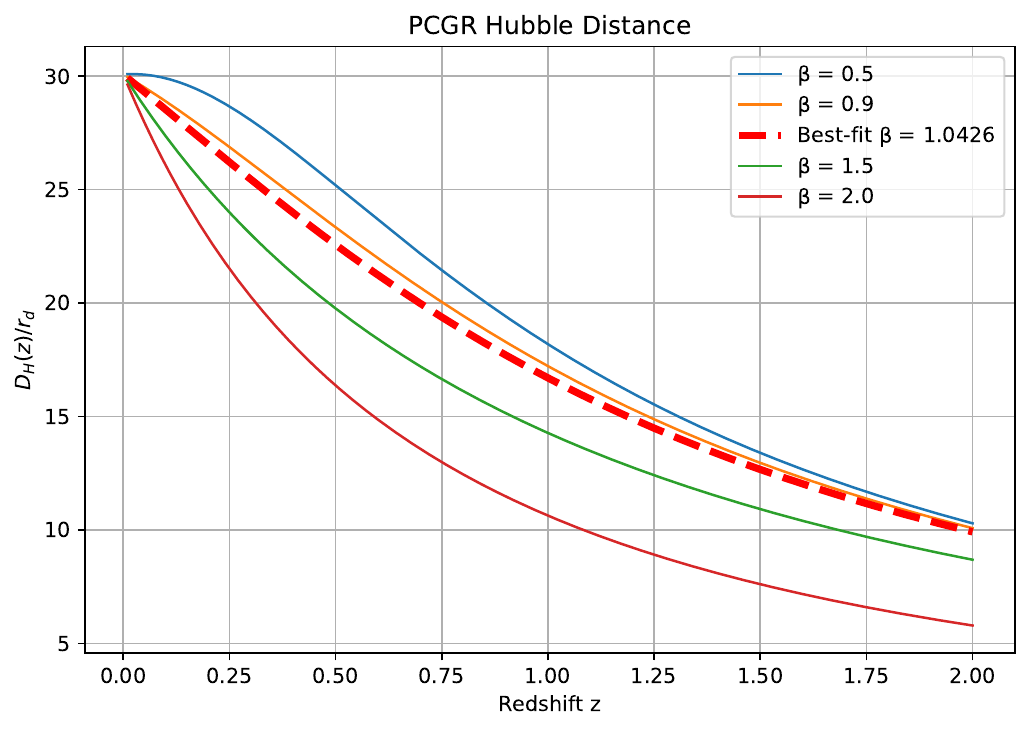}
\caption{
PCGR prediction for the Hubble distance $D_H(z)/r_d$ for
  different values of $\beta$. The broad dashed curve
  corresponds to the best-fit value $\beta = 1.0426$.
	}
\label{Fig:1a}
   \end{minipage}\hfill
   \begin{minipage}{0.48\textwidth}
     \centering
     \includegraphics[width=1\linewidth]{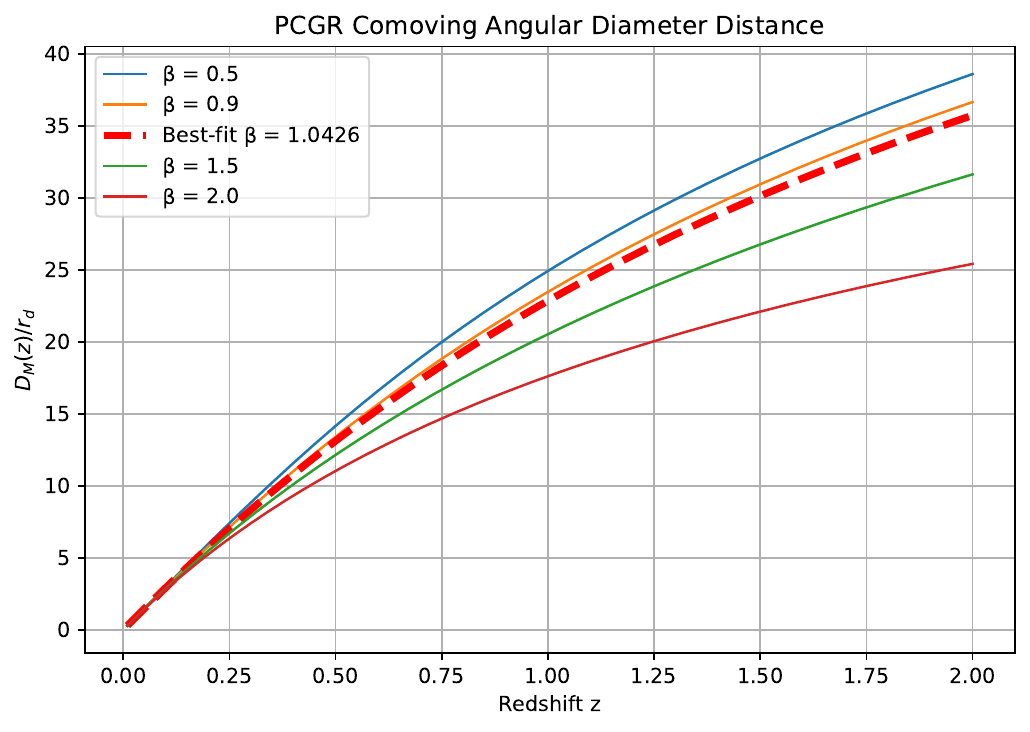}
     \caption{
		PCGR prediction for the comoving angular
         diameter distance $D_M(z)/r_d$ for different values of
         $\beta$. The broad dashed curve indicates the best-fit value
         $\beta = 1.0426$.
				}
	\label{Fig:1b}
   \end{minipage}
\end{figure}

\begin{figure}
\centering \includegraphics[width=0.50\textwidth]{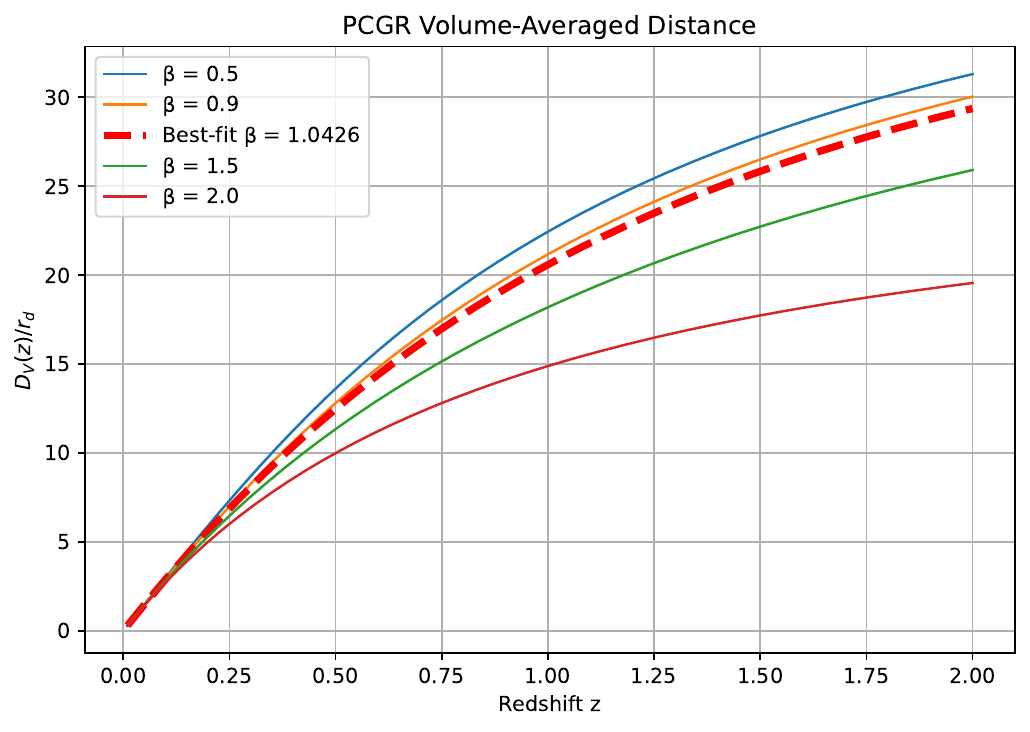}
\caption{
PCGR prediction for the volume-averaged distance $D_V(z)/r_d$
as a function of redshift for different values of the parameter
$\beta$. The broad dashed curve corresponds to the best-fit value
$\beta = 1.0426$.
}
\label{Fig:1c}
\end{figure}

	To illustrate how the pseudo-complex
  deformation parameter $\beta$ affects different BAO distance
  measures, we plot in Figs.~\ref{Fig:1a} to \ref{Fig:1c} the
  pcGR predictions for several values of the parameter $\beta$ and compare them to
  the DESI observables. The three figures correspond to distinct
  cosmological distance measures: the Hubble distance $D_H(z)/r_d$,
  the comoving angular diameter distance $D_M(z)/r_d$, and the
  volume-averaged distance $D_V(z)/r_d$. Each of these quantities
  probes the expansion history in a different way, with $D_H$ being
  directly sensitive to the local expansion rate $H(z)$, while $D_M$
  and $D_V$ encode integrated information along the line of sight. The
  broad dashed curve in each panel indicates the best-fit value $\beta
  = 1.0426$ obtained from the full DESI BAO likelihood analysis.

In the next section, we will explain how to use a statistical method
to compare our theoretical cosmological
to the DESI data.

\subsection{DESI BAO Data and Fitting the pcGR Parameter $\beta$}

The Dark Energy Spectroscopic Instrument (DESI) is currently the most 
precise experiment, mapping the large-scale distribution of galaxies and 
quasars. One of its principal cosmological outputs is the measurement 
of baryon acoustic oscillations (BAO), which provide a standard ruler 
to probe the expansion history of the Universe. Since the 
pcGR model predicts a modified 
Hubble expansion through the parameter $\beta$, DESI BAO observations 
offer a powerful means of constraining the model.

This section presents a detailed description of the DESI BAO data, the 
structure of the mean vector and covariance matrix, and the statistical 
method used to extract the best-fit value of the pcGR parameter $\beta$.

\subsubsection{BAO Observables Measured by DESI}

BAO corresponds to the imprint of acoustic waves in the pre-recombination 
photon--baryon plasma.
This imprint establishes a characteristic comoving scale, the sound 
horizon at the drag epoch $r_d= 147.33~{\rm Mpc}$.  
DESI measures cosmological distances relative to this ruler, typically 
reported as i) $\frac{D_M(z)}{r_d}$, ii) $\frac{D_H(z)}{r_d}$ and 
iii) $\frac{D_V(z)}{r_d}$.

For the dataset used in this work, DESI provides a total of 
thirteen correlated measurements:
\begin{itemize}
    \item one value of $D_V/r_d$,
    \item six values of $D_M/r_d$,
    \item six values of $D_H/r_d$.
\end{itemize}
These measurements form a single coherent data set and must be treated 
together in the likelihood analysis.

\subsubsection{DESI Mean Vector}
The DESI mean vector collects all thirteen measurements into a single 
column vector:
\begin{equation}
y_{\rm data} =
\begin{pmatrix}
(D_V/r_d)(z_1) \\
(D_M/r_d)(z_2) \\
(D_H/r_d)(z_2) \\
\vdots \\
(D_M/r_d)(z_7) \\
(D_H/r_d)(z_7)
\end{pmatrix}.
\end{equation}

The ordering of the components follows the DESI publication and must be 
preserved when constructing the theoretical model vector.  
Each entry corresponds to a specific redshift and a specific BAO 
observable, making this structure essential for a proper likelihood 
evaluation.
From \citet{DESI2024}, we obtain the DESI mean vector dataset, which are listed in
Table \ref{tab1}
\begin{table}
\centering
\begin{tabular}{ccc}
\hline
\textbf{Redshift} $z$ & \textbf{Quantity} & \textbf{Value} \\
\hline
0.295 & $D_V(z)/r_d$ & 7.9417 \\
0.510 & $D_M(z)/r_d$ & 13.5876 \\
0.510 & $D_H(z)/r_d$ & 21.8629 \\
0.706 & $D_M(z)/r_d$ & 17.3507 \\
0.706 & $D_H(z)/r_d$ & 19.4553 \\
0.934 & $D_M(z)/r_d$ & 21.5756 \\
0.934 & $D_H(z)/r_d$ & 17.6415 \\
1.321 & $D_M(z)/r_d$ & 27.6009 \\
1.321 & $D_H(z)/r_d$ & 14.1760 \\
1.484 & $D_M(z)/r_d$ & 30.5119 \\
1.484 & $D_H(z)/r_d$ & 12.8170 \\
2.330 & $D_H(z)/r_d$ & 8.6315 \\
2.330 & $D_M(z)/r_d$ & 38.9890 \\
\hline
\end{tabular}
\caption{DESI DR2 BAO measurements used in this work.}
\label{tab1}
\end{table} 

\begin{figure}
\centering 
\includegraphics[width=0.50\textwidth]{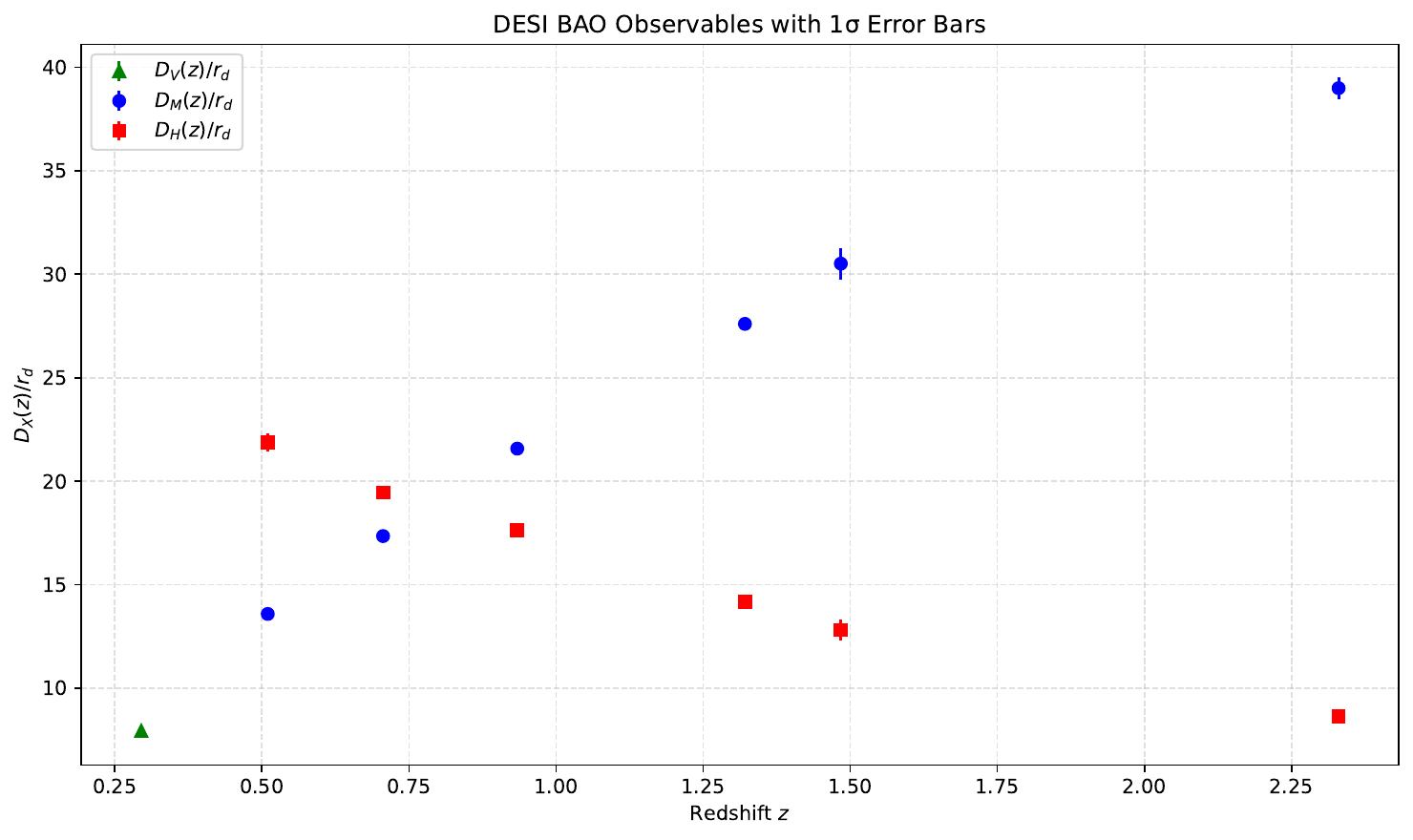} 
\caption{
DESI BAO observables with $1\,\sigma$ error bars, constructed
using the provided mean vector and covariance matrix, demonstrating the viability of the pcGR model given current data.
}
\label{fig2}
\end{figure}

From the DESI mean vector dataset, we obtain the plot depicted in Fig.~\ref{fig2},
which provides a visual summary of the DESI BAO measurements and their
uncertainties, illustrating the relative constraining power of the
different distance observables before comparison with theoretical
predictions.

\subsubsection{DESI Covariance Matrix}

Because the BAO measurements at different redshifts are not statistically 
independent, DESI provides a full $13\times 13$ covariance matrix $C$.  
The diagonal elements contain the variances:
\begin{equation}
C_{ii} = \sigma_i^2,
\end{equation}
while the off-diagonal terms quantify the correlations:
\begin{equation}
C_{ij} \neq 0 \qquad (i \ne j).
\end{equation}

These correlations arise from the fact that different distance 
measurements at the same redshift are extracted from the same galaxy 
sample, and that different redshift bins partially overlap in volume.  
Therefore, a statistically correct comparison with theory requires the 
use of the full covariance matrix rather than treating the measurements 
independently.
From \citet{DESI2024}, we obtain the DESI covariance matrix dataset, whose non-zero entries
are listed in Table \ref{tab2}.

\subsubsection{Likelihood and Estimation of $\beta$}

To determine the value of $\beta$ that best matches the DESI data, the 
standard multivariate Gaussian likelihood is used.  
The corresponding $\chi^2$-
 statistic is

\begin{equation}
\chi^2(\beta)
=
\big( y_{\rm data} - y_{\rm model}(\beta) \big)^{\rm T}\,
C^{-1}\,
\big( y_{\rm data} - y_{\rm model}(\beta) \big).
\end{equation}

This expression accounts simultaneously for all thirteen BAO measurements 
and for the correlations encoded in the covariance matrix.  
It determines how well a given value of $\beta$ reproduces the DESI 
observations.

The best-fit parameter is obtained by minimizing $\chi^2(\beta)$:
\begin{equation}
\beta_{\rm best} = 
\underset{\beta}{\arg\min}\;\chi^2(\beta).
\end{equation}

Because $\beta$ is a single free parameter, the $1\sigma$ confidence interval for a single parameter is obtained from 
\begin{equation}
\chi^2(\beta) - \chi^2_{\rm min} = 1.
\end{equation}

A reduced $\chi^2$ is also computed:
\begin{equation}
\chi^2_\nu =
\frac{\chi^2}{N_{\rm data} - N_{\rm parameters}}
=
\frac{\chi^2}{13 - 1},
\end{equation}
providing a measure of the goodness of fit.
Values of $\chi^2_\nu \approx 1$ indicate an excellent fit,
$\chi^2_\nu \le 2 $ are acceptable, and higher values signal 
significant tension between the model and the data.

\begin{table}
\centering
\begin{tabular}{ll}
\hline
$(i,j)$ & $C_{ij}$ \\
\hline
$(1,1)$ & $5.790\times 10^{-3}$  \\
$(2,2)$ & $2.835\times 10^{-2}$ \\
$(2,1)$ & $-3.261\times 10^{-2}$  \\
$(3,2)$ &  $-3.261\times 10^{-2}$ \\
$(3,3)$ & $1.839\times 10^{-1}$  \\
$(4,4)$ & $3.238\times 10^{-2}$  \\
$(4,5)$ & $-2.374\times 10^{-2}$ \\  
$(5,5)$ & $1.115\times 10^{-1}$ \\
$(6,6)$ & $2.617\times 10^{-2}$ \\
$(6,7)$ & $-1.129\times 10^{-2}$ \\
$(7,6)$ & $-1.129\times 10^{-2}$ \\
$(7,7)$ & $4.042\times 10^{-2}$ \\
$(8,8)$ & $1.053\times 10^{-1}$ \\
$(8,9)$ & $-2.903\times 10^{-2}$  \\
$(9,8)$ & $-2.903\times 10^{-2}$ \\
$(9,9)$ & $5.042\times 10^{-2}$ \\
$(10,10)$ & $5.830\times 10^{-1}$ \\
$(10,11)$ & $-1.952\times 10^{-1}$ \\
$(11,10)$ & $-1.952\times 10^{-1}$ \\
$(11,11)$ & $2.683\times 10^{-1}$ \\
$(12,12)$ & $1.021\times 10^{-2}$ \\
$(12,13)$ & $-2.314\times 10^{-2}$ \\
$(13,12)$ & $-2.314\times 10^{-2}$ \\
$(13,13)$ & $2.827\times 10^{-1}$ \\
\hline
\end{tabular}
\caption{Non-zero elements of the covariance matrix.
}
\label{tab2}
\end{table}

\begin{figure}
\centering 
\includegraphics[width=0.50\textwidth]{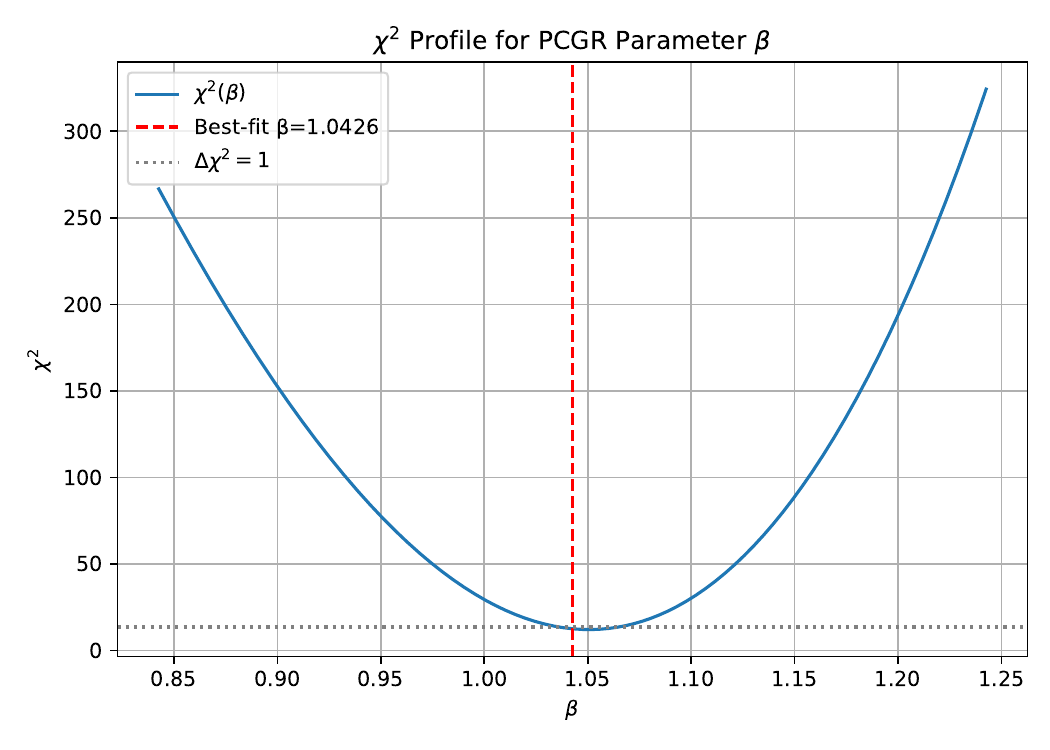} 
\caption{
The dependence of the $\chi^2$ function on the pcGR parameter
$\beta$. The minimum at $\beta = 1.0426$ identifies the best-fit value
from the DESI BAO likelihood analysis.
}
\label{fig3}
\end{figure}

This statistical procedure allows for a rigorous and unbiased comparison 
between the pcGR theoretical model and DESI BAO observations, enabling a 
precise determination of  the parameter 
$\beta$ and a direct test of the viability of the pcGR cosmology.
In order to assess the quality of the fit for $\beta$, we examine the
dependence of $\chi^2(\beta)$ on $\beta$, as shown in
Fig.~\ref{fig3}.
The smooth, approximately parabolic shape of the
$\chi^2(\beta)$ curve, together with the presence of a single,
well-defined minimum, indicates the absence of degeneracies in the fit
and demonstrates that the parameter $\beta$ is tightly and robustly
constrained by the DESI BAO data.

From the statistical analysis and the combined fit of the DESI
observables, we obtain the best-fit value of the pcGR parameter
$\beta$. By minimizing the $\chi^2$ function constructed from the DESI
mean vector and its covariance matrix, the optimal value is found to
be $\beta = 1.04261$.

\begin{figure}
\centering 
\includegraphics[width=0.40\textwidth]{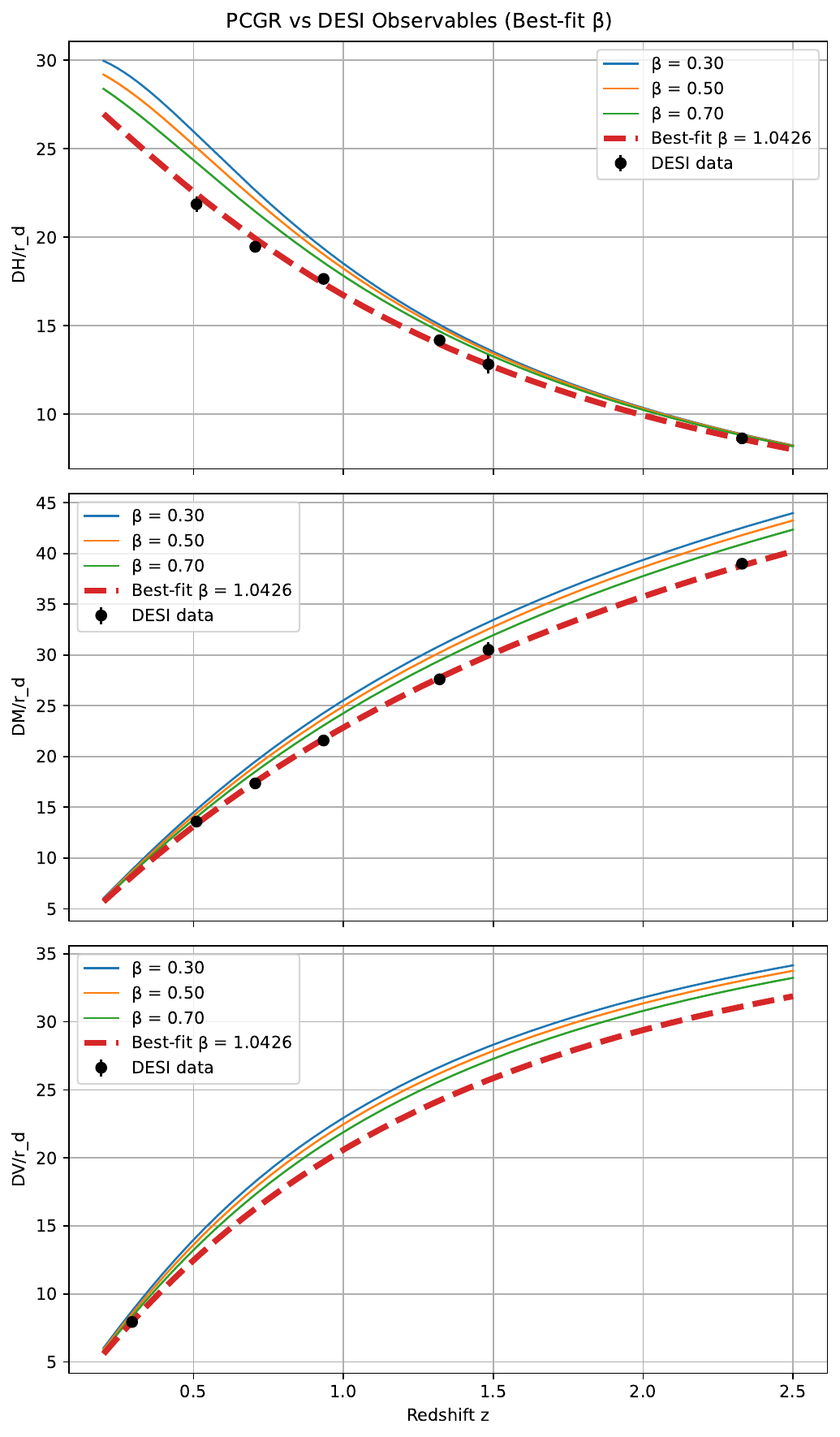}
\caption{
DESI 2024 BAO measurements of $D_H(z)/r_d$, $D_M(z)/r_d$, and
$D_V(z)/r_d$ compared with pcGR predictions for several values of
$\beta$. The broad dashed curve corresponds to the best-fit value
$\beta = 1.04261$ and shows excellent agreement with all DESI
  observables simultaneously, demonstrating the viability of the model.
	}
\label{fig4}
\end{figure}

This result represents the value of $\beta$ that provides the closest
agreement between the theoretical predictions of pseudo-complex
General Relativity and the current DESI observational data.  The plots
of the cosmological observables $D_{H}(z)/r_{d}$, $D_{M}(z)/r_{d}$,
and $D_{V}(z)/r_{d}$ compare the DESI BAO measurements with their
$1\sigma$ uncertainties
to the pcGR predictions for several values of the parameter $\beta$ 
(see Figs.~\ref{Fig:1a}, \ref{Fig:1b}, and \ref{Fig:1c}). 
These curves show how sensitively each observable responds to variations in $\beta$, allowing a direct visual assessment of the model’s behavior across redshift. Figure~\ref{fig4} shows that by computing the 
$\chi^2$ function $\chi^{2}(\beta)$ using the full DESI covariance matrix, we identify a clear minimum at the best-fit value 
$\beta=1.04261$. 
This $\beta$-value implies that $w_\Lambda = -1.0426$, slightly less
than $-1$. This places the pcGR effective fluid in the ``phantom''
domain ($w < -1$) \citet{Caldwell2003}, but \textit{critically}, within
pcGR this arises from geometry, not an exotic matter field, and
without the usual instabilities appearing in a phantom field.
It is important to emphasize that the parameter $\beta$ is not a free
phenomenological deformation of the dark-energy equation of state.
Instead, $\beta$ originates purely from the geometric constraint
$g_{\mu\nu} u^{\mu} u^{\nu} = C$ that follows from the pseudo-complex
structure of the theory.  Because the deviation $\beta - 1$ is tied
directly to the relation $\beta = 1 +
\frac{2}{3}\frac{\dot{H}}{H^{2}}$, the effective value $w_\Lambda =
-\beta < -1$ does not correspond to a matter field with a negative
kinetic term.  Consequently, the model avoids the ghost and gradient
instabilities that typically plague phantom scalar-field theories,
since no additional dynamical degree of freedom is introduced and the
modification resides entirely in the geometric sector.

Using the DESI BAO mean vector and covariance matrix, we perform a
one–parameter fit of the pseudo–complex cosmological model. The minimization
of the $\chi^{2}$ function yields a best--fit value for
$\beta_{\rm best}$, 
with a corresponding minimum
$\chi^{2}_{\min} = 12.633$ ,
$\chi^{2}_{\nu} = \frac{\chi^{2}_{\min}}{\nu} = 1.053$
for $\nu = 12$ degrees of freedom.
Since $\chi^{2}_{\nu} \approx 1$, the PCGR predictions are statistically
consistent with the DESI BAO data and provide an excellent fit.
Using the standard $\Delta \chi^{2} = 1$ criterion for one parameter, we obtain
the $1\sigma$ confidence interval
$\beta = 1.0426 \pm 0.0144$ 
(with the error defined as $\delta \beta=0.0144$),
showing that the deformation parameter $\beta$ is tightly constrained.
The smooth parabolic behavior of $\chi^{2}(\beta)$ further confirms the
robustness and stability of the fit.
In \citet{Planck2020} a combined analysis of Planck-BAO-SNIa data is presented, with a 
value of $w_\Lambda = -1.028 \pm0.032$, which is consistent with our estimate.

The $\chi^2$ value 
obtained
is just above 1, implying an excellent fit.  This point represents the
theoretical curve that most closely follows the DESI data in all three
observables simultaneously. The good agreement between the DESI points
and the pcGR predictions at this value provides a strong verification
of the model within current observational precision. Overall, the
plots demonstrate that pcGR remains consistent with modern BAO
constraints while allowing controlled deviations from $\Lambda$CDM
through the single parameter $\beta$.

\subsection{Consequences}

In what follows, we deduce the Hubble acceleration and the
acceleration parameter.  In \citet{Hess2015} [p. 118, Eq. (4.99)] and
in Eq. (\ref{sec2-4}) above, the relation between $\beta$, the Hubble
constant, and the time derivative of the Hubble constant is given as
$\beta  =  1 + \frac{2}{3}\frac{\dot{H}_0}{H_0^2}$,
where the dot refers to the derivative in time.
This equation implies that $\beta \ge 1$, i.e., the difference to 1 gives information about
the time dependence of the Hubble "constant".
Solving (\ref{sec2-4}) for $\dot{H}$ gives
\beqa
\dot{H}_0 & = & (\beta - 1)\frac{3}{2} H_0^2
~~~.
\label{intro-7}
\eeqa
For numerical evaluation, we adopt a present-day Hubble constant
$H_0 = 67.4~\mathrm{(km/s) / Mpc}$, consistent with current
Planck and DESI analyses. While the numerical value of $\dot H_0$
scales with the assumed $H_0$, its order of magnitude is unchanged.
Converting to SI units using
$1~\mathrm{Mpc} = 3.086\times10^{19}~\mathrm{km}$, this corresponds to
\beqa
H_0 = 67.4 ~ \mathrm{(km/s)/Mpc} =  2.18\times 10^{-18}~\mathrm{s^{-1}} ~~~ .
\label{intro-9}
\eeqa
Using the best-fit DESI value $\beta = 1.04261$, Eq.~(\ref{intro-7}) then yields
\beqa
\dot H_0 &=& 0.04261 \, \frac{3}{2} \, H_0^2 ~~~.
\label{intro-8}
\eeqa
Substituting the SI value of $H_0$ from Eq.~(\ref{intro-9}) into
Eq.~(\ref{intro-8}), we obtain
\beqa \dot{H}_0 & \approx & (0.9388 \pm
0.317) \times 10^{-17} ~ \mathrm{(km/s^2)/Mpc} \\
&\approx& (0.3012 \pm
0.102) \times 10^{-36}~\mathrm{s}^{-2}
~~~,
\label{intro-10}
\eeqa
where we have determined the error propagation in the Appendix A.
One has to keep in mind that when another value for $H_0$ is used, as in (\ref{intro-9}),
this value changes in scale accordingly.


  
This can be related to the standard deceleration parameter $q$
(historically referred to as the deceleration parameter in
\citet{Adler1975,Misner1973}), which is defined as

\begin{eqnarray}
  q & \equiv & -\frac{\ddot{a}a}{\dot{a}^2} ~~~.
  \label{eq:q.decel}
\end{eqnarray}

In the present work, we adopt the relation

\begin{eqnarray}
\dot{H}_0 &=& H_0^2 \left( 1 + q \right) ~~~,
\label{intro-11}
\end{eqnarray}
which corresponds to the convention that $\dot{H}_0 = 0$ implies
$q = -1$.
Combining this expression with Eq.~(\ref{intro-7}), we obtain
\begin{eqnarray}
q &=& \frac{3}{2}\left(\beta - \frac{5}{3}\right) ~~~.
\label{intro-12}
\end{eqnarray}

With $\beta = 1.04261$, we obtain $q=-0.93609 \pm 0.0216$. This value
is significantly more negative than the $\Lambda$CDM prediction of
$q_{\Lambda\mathrm{CDM}} \approx -0.55$ for the same $\Omega_m$,
indicating that the expansion of the Universe is accelerating more
rapidly in the pcGR model. The deceleration parameter $q$
[Eq.~(\ref{eq:q.decel})] quantifies the second derivative of the scale
factor. A more negative value of $q$ corresponds to a greater
acceleration relative to the Hubble expansion rate.

These obtained values allow us to determine the {\it
  redshift drift} or {\it Sandage-Loeb effect} \citet{Lazkoz},
calculated in the 
Appendix B
Using the exact Sandage--Loeb relation and the full model-dependent Hubble
function $H(z)$, we find a redshift drift of
$\Delta v \simeq -11.1\ \mathrm{cm\, s^{-1}}$ over 20 years at $z=4$ for the
best-fit pcGR model. The corresponding $\Lambda$CDM prediction is
$\Delta v \simeq -11.0\ \mathrm{cm\, s^{-1}}$, implying a difference of order
$0.1\ \mathrm{cm\,s^{-1}}$ over the same interval.
Scaling linearly with time, this corresponds to
$\sim 5.5\ \mathrm{cm\,s^{-1}}$ per decade.


\section{Conclusions}
\label{sec4}

A basic introduction to the pcGR was given, with the intent to apply it to observations by
DESI. The motivation was to deduce from the data the acceleration of the Universe,
within the pcGR.

The cosmological predictions of the pseudo-complex General Relativity
(pcGR) are compared to observational data from the Dark Energy
Spectroscopic Instrument(DESI).  The pcGR introduces a natural
geometric extension of General Relativity, which modifies the FLRW
equations and leads to an effective dynamical description of the dark
energy, characterized within pcGR by a parameter $\beta$. This
parameter governs the evolution of the dark energy density and relates
to the effective equation of state via $p_\Lambda = -\beta
\varepsilon_\Lambda$ = $w_\Lambda \varepsilon_\Lambda$.
Because the dark energy has within the FLRW model a geometric origin, it represents an
alternative way to explain its emergence.  

The parameter $\beta$ was adjusted to the DESI data, resulting in a
value of $1.04261 \pm 0.0144$.  
This corresponds to
$w_\Lambda = -1.04261 \pm 0.0144$, which can be also obtained in a
model with a phantom energy present \citet{Caldwell2003}, while in pcGR
the origin is completely of geometric nature.  The obtained
$w_\Lambda$ value was also compared to a combined analysis of
Planck-BAO-SNIa \citet{Planck2020}, which is $w_\Lambda = -1.028 \pm
0.032$, which is consistent with our estimate.  
From
there, the Hubble acceleration was deduced, giving a value of 
$\dot{H}_0
= (0.9388 \pm 0.317) \times 10^{-17} ~ \mathrm{(km/s^2)/Mpc}$, 
or $(0.3012 \pm 0.102) \times
10^{-36} ~ {\rm s}^{-2}$. 
From this we to obtain the
acceleration parameter 
$q=-0.93609 \pm 0.0216$.

The derived value of $\dot H_0$ corresponds to a spectroscopic redshift
drift of $\Delta v \simeq -11.1\ \mathrm{cm\,s^{-1}}$ over 20 years for a
source at $z=4$, as obtained from the exact Sandage--Loeb relation.
While the absolute redshift-drift signal is within the reach of future
high-precision spectroscopic facilities, the small difference between the
pcGR and $\Lambda$CDM predictions highlights the need for extremely precise
measurements of the expansion history to directly constrain ${\dot H}_0$.

These results transform $\dot{H}_0$ from an abstract derivative into a
concrete observational target for next-generation facilities such as
the ANDES spectrograph on the ELT \citet{Alves2019ANDES,Martins2016ANDES}
or the SKA \citet{Klockner2015SKA}. This elevates $\dot H_0$ from a fitted
quantity to a concrete, testable prediction for the coming decade.

It is important to emphasize that BAO measurements constrain the pcGR
parameter $\beta$ indirectly through its impact on the integrated
expansion history, whereas a measurement of the redshift drift provides
a direct and model-dependent probe of the time derivative of the Hubble
parameter, $\dot H$. Calculations show that ANDES could reach a
precision of $\sim 1\%$ on $\dot{H}_0/H_0^2$ over a decade \citet{Liske}.


In summary, we have demonstrated that the pseudo-complex extension of
GR provides a consistent cosmological framework that naturally
incorporates a dynamic component to dark energy. Constrained by
state-of-the-art DESI BAO data, the model yields a best-fit value for
$\beta$.  This prediction distinguishes pcGR from the $\Lambda$CDM
baseline ($\dot{H}_0=0$) and aligns it with phenomenological models
favoring a slightly phantom equation of state, while offering a
distinct geometric origin. Future precision measurements of the
expansion history, particularly those sensitive to temporal changes
like redshift drift, will provide a decisive test of this prediction
and the pcGR framework.

\section*{Acknowledgments}
L.M. and P.O.H. acknowledge financial support from PAPIIT-DGAPA (IN116824).



\bibliographystyle{mnras}
\bibliography{Hess} 

@article{Planck2020,
author={Planck Collaboration,\textit{Planck 2018 results. VI. Cosmological parameters}},
title={Not a Real Paper},
journal={Astron.~Astrophys.},
volume=641,
pages={A6},
year=2018,
}

@article{Hogg.1999,
author={Hogg, D. W.},
title={Not a Real Paper},
journal={arXiv pfeprint, astro-ph/9905116},
year=1999}

@article{HubbleSNIa,
author={Bora,  K. and Holanda, R. F. L.},
journal={Eur. Phys. J.},
volume=83,
pages={274},
year=2023
}

@article{Riess2022,
author={Riess, A. G. and Yuan, W. and  Macri, L. M. and  et al.},
journal={The Astr. Jour. Letters},
volume=934,
pages={L7},
year=2022
}

@article{DESI2024,
author={Abdul-Karim, M. and  Aguilar, J. and Ahlen, S. and et al.},
journal={Jour. Cosm. and Astropart. Physics},
volume=04,
pages={012},
year=2025
}

@article{DESI2025a,
author={Lodha, K. and Calderos R. and Mathewson, W. L. and et al.},
journal={Phys. Rev. D},
volume=122,
pages={1083512},
year=2025
}

@article{DESI2025b,
author={Andrade, U. and Paillas, E. and  Mena-Fern\'andez, J. and et al.},
journal={Phys. Rev. D},
volume=122,
pages={1083512},
year=2025
}

@article{DESI-data,
author={Abdul-Karim, M. and DESI-Collaboratio},
journal={Phys. Rev. D},
volume=122,
pages={083515},
year=2025
}

@article{Copeland,
author={Copeland, E. J. and  Sami, M. and  Tsujikawa, S.},
journal={International Journal of Modern Physics D},
volume=15,
pages={573},
year=2006
}

@article{Hess2020,
author={Hess, P. O.},
journal={Prog. in Part. and Nucl. Phys.},
volume=114,
pages={103809},
year=2020
}

@article{Kelly1986,
author={Kelly, P.F. and  Mann, R. },
journal={Class. and Quant. Grav.},
volume=3,
pages={705},
year=1986
}

@article{Leila2025,
author={Maghlaoui, L. and Hess, P. O.},
journal={General Relativity and Gravitation},
volume=57,
pages={133},
year=2025
}

@article{Weber2025,
author={Weber, F. and  Hess, P. O. and Zen Vasconcellos, C. A.},
journal={Physica Scripta},
volume=100,
pages={095301},
year=2025
}

@article{Leila2010,
author={Hess, P. O. and Maghlaouil, L.},
journal={Int. J. of Mod. Phys. E},
volume=19,
pages={1315},
year=2010
}

@article{Caldwell2003,
author={Caldwell, R. R. and  Kamionkowski, M. and Weinberg, N. N.},
journal={Phys. Rev. Lett.},
volume=91,
pages={071301},
year=2003
}

@article{Alves2019ANDES,
author={Alves, C. S. and Martins, C. A. J. P. and Charnock, T.},
journal={MNRAS},
volume=488,
pages={3607},
year=2019
}

@article{Lazkoz,
author={Lazkoz, R. and Leanizbarrutia, I. and  Salzano, V.},
journal={The European Physical Journal C},
volume=78,
pages={},
year=2018
}

@article{Martins2016ANDES,
author={Martins, C. J. A. P. and  et al.},
journal={arXiv},
pages={1605.08761,1605.08761},
year=2016
}

@article{Klockner2015SKA,
author={Klockner, H.-R. and et al.},
journal={Proceedings of Science},
volume=215,
pages={027},
year=2015
}

@article{Liske,
author={Liske, J. and et al.},
journal={MNRAS},
volume=386,
pages={1192},
year=2008
}

@book{Adler1975,
  author       = {Adler, R. and  Bazin, M. and  Shiffer, M.},
  title        = {Introduction to General Relativity},
  publisher    = {Mc Graw Hill}, 
year    ={1975},
 isbn  = {10.0070004234}
}

@book{Hess2015,
  author       = {Hess, P. O. and Schäfer, M. and Greiner, W.},
  title        = {Pseudo-Complex General Relativity},
  publisher    = {Springer},
  year         = {2015},
  series       = {FIAS Interdisciplinary Science Series},
  volume       = {7366},
 doi          = {10.1007/978-3-319-25061-8},
  isbn         = {978-3-319-25060-1},  
  url          = {https://doi.org/10.1007/978-3-319-25061-8}
}

@book{Misner1973,
  author       = {Misner, C. W. and Thorne, K. S. and  Wheeler, J. A.},
  title        = {GRAVITATION},
  publisher    = {W. H. Freeman and Company},
  year         = {1973},
  isbn         = {0-7161-0334-3}, 
}




\appendix
\section*{Appendix A: Determination of the error propagation of ${\dot H}_0$
and $q$}

Including the error $\delta \beta = 0.0144$ in $\beta$, we have for the
time derivative of the Hubble constant 
\beqa \dot{H}_0 + \delta \dot{H}_0  =  \frac{3}{2} H_0^2 \left(
(\beta - 1) + \delta \beta \right)
\eeqa
 with 
 \beqa
 \delta \dot{H}_0  =  \frac{3}{2} H_0^2 \delta \beta ~=~
\left[ \frac{3}{2} H_0^2 (\beta -1) \right] \frac{\delta \beta}{(\beta
  - 1)} ~~~.
\label{apb-1}
\eeqa
The expression in square brackets evaluates to
$0.9388 \times 10^{-17}\,(\mathrm{km/s^2})/\mathrm{Mpc}$.
 Using $(\beta -
1)=0.04261$ and $\delta \beta = 0.0144$, we arrive at
\beqa
\delta \dot{H}_0 & = & 0.3173 \times 10^{-17} ~ \mathrm{(km/s^2)/Mpc}
~~~ .
\eeqa 

For the acceleration parameter the information on its error is directly obtained, noting
that $q+\delta q=\frac{3}{2}\left( (\beta + \delta\beta) - \frac{5}{3}\right)$
and $\delta\beta = 0.0144$, which results in
\beqa
\delta q & = & \frac{3}{2}\delta\beta ~=~ 0.0216
~~~.
\eeqa

\section*{Appendix B: Calculation of the Predicted Redshift Drift}\label{app:redshift_drift}

This appendix details the calculation connecting the cosmological expansion
history predicted by the pcGR model to a measurable spectroscopic velocity
shift (redshift drift), also known as the Sandage--Loeb effect.

\vskip 0.5cm
\noindent
\textit{Theoretical Foundation}

The cosmological redshift drift quantifies the change in the observed redshift
of a distant source due to the time evolution of the cosmic expansion rate.
For a source at redshift $z$, observed over a time interval $\Delta t_0$ at the
present epoch, the exact expression for the spectroscopic velocity shift is
given by \citet{Lazkoz}:
\begin{equation}
\Delta v(z) \;=\; c\,\Delta t_0
\left[
H_0 - \frac{H(z)}{1+z}
\right] ~ ,
\label{eq:SL_exact}
\end{equation}
where $c$ is the speed of light, $H_0$ is the present-day Hubble constant, and
$H(z)$ is the Hubble parameter evaluated at the redshift of the source.
Equation~\eqref{eq:SL_exact} is exact and does not rely on any low-redshift or
Taylor expansion. For sources at high redshift (e.g.\ $z \ge 1$), this expression
must be used directly; approximate first-order expansions around $z=0$ are not
reliable in this regime.

\vskip 0.5cm
\goodbreak
\noindent
\textit{Hubble Function in the pcGR Model}

In the pseudo-complex General Relativity (pcGR) framework, the Hubble parameter
as a function of redshift is given by Eq.~(\ref{H-beta}) of the main text:
\begin{equation}
H^2(z) = H_0^2 \left[
\Omega_m (1+z)^3
+ \Omega_r (1+z)^4
+ \Omega_\Lambda (1+z)^{3(\beta-1)}
\right] ~ ,
\label{eq:Hz_pcgr}
\end{equation}
where $\beta$ is the deformation parameter determined from the DESI BAO fit.
For $\beta=1$, this expression reduces to the standard $\Lambda$CDM form.

\vskip 0.5cm
\noindent
\textit{Numerical Inputs}

For the numerical evaluation, we adopt a present-day Hubble constant
$H_0 = 67.4\ \mathrm{km\,s^{-1}\,Mpc^{-1}}$, the best-fit pcGR deformation
parameter $\beta = 1.0426$, an observational baseline of
$\Delta t_0 = 20\ \mathrm{yr}$, and a source redshift $z = 4$.

The corresponding time interval in seconds is
\[
\Delta t_0 = 20 \times 3.15576 \times 10^7 \ \mathrm{s}
= 6.311 \times 10^8 \ \mathrm{s}~ .
\]

Converting the Hubble constant to SI units gives
\[
H_0 = 67.4 ~ \mathrm{(km/s)/Mpc}
= 2.185 \times 10^{-18} \ \mathrm{s^{-1}}~ .
\]

\vskip 0.5cm
\noindent
\textit{Numerical Evaluation at $z=4$}

It is convenient to write Eq.~\eqref{eq:SL_exact} in terms of the dimensionless
expansion rate $E(z) \equiv H(z)/H_0$:
\begin{equation}
\Delta v(z) = c\,H_0\,\Delta t_0
\left[
1 - \frac{E(z)}{1+z}
\right] ~ .
\label{eq:dv_Ez}
\end{equation}
Using representative cosmological parameters
$\Omega_m \simeq 0.315$ and $\Omega_\Lambda \simeq 0.685$ (with the radiation
contribution negligible at $z=4$), the dimensionless expansion rate
$E(z)=H(z)/H_0$ takes the value $E(4)\simeq 6.33$ in the standard
$\Lambda$CDM case ($\beta=1$). For the pcGR model with the best-fit
parameter $\beta=1.0426$, the corresponding value is only slightly larger,
$E(4)\simeq 6.34$, reflecting the small deviation from $\Lambda$CDM at this
redshift.

The overall prefactor in Eq.~\eqref{eq:dv_Ez} evaluates to
\[
c\,H_0\,\Delta t_0
= (3\times10^8\ \mathrm{m\,s^{-1}})
(2.185\times10^{-18}\ \mathrm{s^{-1}})
(6.311\times10^8\ \mathrm{s})
\simeq 0.414\ \mathrm{m\,s^{-1}}~ .
\]

Substituting these values into Eq.~\eqref{eq:dv_Ez}, the predicted redshift
drift at $z=4$ over a 20-year observational baseline is
\begin{align}
\Delta v_{\Lambda\mathrm{CDM}}(4)
&\simeq 0.414
\left(1 - \frac{6.33}{5}\right)
\simeq -0.110\ \mathrm{m\,s^{-1}}
= -11.0\ \mathrm{cm\,s^{-1}}~ , \\
\Delta v_{\mathrm{pcGR}}(4)
&\simeq 0.414
\left(1 - \frac{6.34}{5}\right)
\simeq -0.111\ \mathrm{m\,s^{-1}}
= -11.1\ \mathrm{cm\,s^{-1}}~ .
\end{align}

The negative sign indicates a blueshift, which is the expected
signature for sources at high redshift in an accelerating
Universe. The difference between pcGR and $\Lambda$CDM at $z=4$ is of
order $0.1\ \mathrm{cm\,s^{-1}}$ over a 20-year baseline. This
calculation directly connects the pcGR expansion history to a
measurable spectroscopic redshift drift at high redshift.

\bsp	
\label{lastpage}
\end{document}